\documentclass{article}
\usepackage{graphicx}
\usepackage[margin=1in]{geometry}
\usepackage[T1]{fontenc}
\usepackage[utf8]{inputenc}
\usepackage{lmodern}
\usepackage{microtype}
\usepackage{booktabs}
\usepackage{array}
\usepackage{calc}
\usepackage{caption}
\usepackage{amsmath}
\usepackage{tabularx}
\usepackage{hyperref}
\newcommand\tightlist{\setlength{\itemsep}{0pt}\setlength{\parskip}{0pt}}
\hypersetup{
  pdftitle={Growing One's AI Tools: A Paradigm of Practice-Grown Tools},
  pdfauthor={Zhao Zeming},
  hidelinks}

\title{Growing One's AI Tools: A Paradigm of Practice-Grown Tools}
\author{Zhao Zeming}
\date{}

\begin{document}
\maketitle

\textbf{Abstract}

Professionals have long used software built for them by others. This article reports the growth of a professional computer-assisted interpreting (CAI) tool built by its own user, a conference interpreter without programming training. Drawing on an eight-week version-controlled record with telemetry from 133 logged sessions and over 20,000 recognized utterances, the article traces how real conference work drove the tool through 62 commits and six versions. The record shows capabilities deployed on the day their need appeared and one strategy change that cut median end-to-end delay by over a third within one afternoon. The article then proposes the paradigm of practice-grown tools, defined by the shift from using to growing, the co-evolution of tool and practice through breakdown and repair, bounded local computation, and an assessment corollary, and it formulates three propositions toward tool-inclusive assessment of practitioners, from tool-making literacy to a compute-standardized format for certifying what they and their tools achieve together.


\section{1. Introduction}

Professional work has long run on software that professionals themselves did not make; the user-innovation literature documented this producer-centered arrangement across many industries (von Hippel, 2005). Human--computer interaction has correspondingly organized itself around the use, the usability, and the acceptance of finished tools (Davis, 1989; Nielsen, 1993), yet a quiet change in that arrangement is now visible at the edges of many professions, as conversational code generation lets practitioners who cannot program describe what they need and receive working implementations of it. When such a practitioner builds working software this way and then keeps growing it as real work exposes new needs, the interesting questions move past whether it can be built at all. The questions become whether the growing can be sustained under production conditions, whether it generalizes into a pattern that other professions would recognize, and what it demands of the institutions that train and certify practitioners.

This study is grounded in one such profession, conference interpreting, in which the interpreter works in the simultaneous mode, rendering continuous speech from one language into another in real time, seconds behind the speaker and before a live audience. Gile's (2009) effort models describe the work as the concurrent management of listening, memory, production, and coordination under acute time pressure, so tool latency or failure has immediate and public consequences for the delivered service. Interpreters have long worked with tools built for them by others, and the computer-assisted interpreting literature has examined such systems from the developers' side (Fantinuoli, 2018); the record analyzed here comes from the interpreter's own side of that division. Throughout the present article, \emph{the practice} refers to this interpreter's real conference work; \emph{the practitioner}, to the interpreter who conducts it, who is also the author; and \emph{the tool}, to the working software that the same interpreter built for that work and has kept growing since.

This paper is the second in a series that follows the tool; a founding study documented how a professional conference interpreter without programming training built a working computer-assisted interpreting tool through co-creation with AI systems, and followed the build from first dialogue to the first fully instrumented assignment (Zhao, 2026a). That study settled the feasibility of the first turn of the loop, and what happened afterward, as the tool entered regular professional service and kept changing under real conference work, is the subject of the present paper.

The evidence base of this study is an eight-week, version-controlled growth record (August 27 to October 22, 2026) covering 62 commits and six tagged versions, together with telemetry from 133 logged sessions amounting to over 20,000 recognized utterances captured during real conference work and its validation. The record shows a tool that responded to practice almost as fast as practice demanded, releasing subtitle upgrades on the day of a challenging four-session conference, deploying a language-identification capability on the day of a six-language relay forum, and cutting its median delay by over a third on the very afternoon that exposed the problem. It also shows the tool passing through a three-day, high-density forum without triggering a single change; the author reads that stability as the moment when an experiment became infrastructure.

On this record the paper proposes the paradigm of practice-grown tools, in which a domain expert originates, specifies, judges, and maintains the software of the expert's own practice while AI systems supply implementation, and in which the tool grows through recurring cycles of breakdown and repair rather than through planned releases. Four properties define the paradigm, covering the shift from using to growing, the co-evolution of tool and practice, computation that is bounded and local by design, and an assessment corollary that Section 5 develops. An emerging ideal within the paradigm is a division of labor in which the expert designs the load-bearing skeleton and AI generation grows the outer layers, and the paradigm is offered as a candidate whose boundary conditions are stated plainly.

If tools now co-evolve with practice, then the assessment of practitioners cannot indefinitely proceed as though the tools did not exist. The paper advances three propositions in this direction, arguing for a tool-making literacy as the successor to tool proficiency, for the practitioner together with the tools actually at hand as the unit of assessment, and for a compute-standardized format in which candidates create tools on identical hardware under identical task conditions.

The paper sets out four contributions, listed in the order of the argument:
\begin{enumerate}
\tightlist
\item The naming and internal structuring of a paradigm that current literature on conversational code generation documents only in fragments.
\item A longitudinal, telemetry-backed growth record of a high-stakes professional tool grown by its own user, deposited as an open dataset alongside the open repository of the tool itself.
\item Three propositions toward tool-inclusive assessment, so far unaddressed by certification regimes in the language professions.
\item The boundaries of the candidate paradigm, specifying where the paradigm applies and where it should not be expected to hold.
\end{enumerate}

The sections that follow take the components of this argument in turn. Section 2 reviews the four literatures that meet at the corner this study occupies, and Sections 3 through 5 then carry the core of the argument, developing the paradigm, presenting the growth record, and advancing the assessment propositions in that order. The paradigm is stated ahead of its evidence, since the growth record is read through the paradigm's vocabulary of breakdown and repair. Section 6 discusses implications, risks, and limitations for the wider field, and Section 7 draws the argument together. Three questions stay deliberately open; they belong to studies that this record can motivate and not answer, namely whether aided performance changes when interpreters work with grown tools, whether practitioners in other domains replicate the growth pattern documented here, and what a compute-standardized assessment format reveals when it is first tried, and the dataset and the repository are openly available for those studies.

\section{2. Related Work}

\subsection{2.1 User Innovation and End-User Development}

Innovation research established over four decades ago that users, and above all lead users who face demanding needs ahead of the general market, frequently develop solutions for themselves and sometimes see those solutions diffuse into entire industries (von Hippel, 1986). The economics of this redistribution were worked out twice. The toolkit studies showed manufacturers shifting design work to users by supplying toolkits for innovation (von Hippel \& Katz, 2002), and later modeling traced the shift from producer innovation to open user collaboration (Baldwin \& von Hippel, 2011; von Hippel, 2005). Within human--computer interaction, the end-user development tradition took up the same ambition at the level of the workstation, arguing that the boundary between using and programming is a designable one and building environments in which practitioners tailor systems to their own needs (Lieberman et al., 2006). An early strand of that tradition recognized that tailoring is itself skilled and social work, carried out by relatively few users on behalf of their colleagues (MacLean et al., 1990), a theme that returns in Section 5. Meta-design condensed the agenda into the image of systems delivered as seeds and grown by their users, with periodic consolidation phases that restructure what use has accumulated (Fischer, 1998; Fischer et al., 2004). Across these literatures, the implementing skill remained scarce, and documented cases of user-developed tools therefore clustered among users with unusual technical resources or access to professional help.

\subsection{2.2 Conversational Code Generation and Vibe Coding}

Conversational code generation has moved the implementation barrier that earlier user-innovation research treated as fixed, and the practice acquired a name in February 2025, when Karpathy (2025) coined the term ``vibe coding'' for programming in which the developer describes intent in natural language and accepts generated code under light review. Empirical work quickly formed around the term and now covers several professional domains. Interview and survey studies characterize the activity as programming through conversation, with effort shifting toward context management, review, and judgment and away from syntax (Sarkar \& Drosos, 2025), and qualitative studies of co-creation document the flow, trust, and communication dynamics between programmers and models (Pimenova et al., 2026). Surveys of the emerging field map its practices and open problems (Ge et al., 2025), design research traces how professionals adopt generation into prototyping work (Kobiella et al., 2026), and a scoping review of healthcare applications finds clinicians already building tools for clinical, administrative, and research tasks (Huynh \& Rivero, 2026). At the community scale, agent-based reverse engineering tooling drew several hundred thousand participants within two months of release, many of them using it to understand and rebuild the software beneath their own workflows (morluto, 2026). Longitudinal evidence from high-stakes professional deployment has been slower to appear in this young literature, and the founding study of the present case contributes exactly that, following one interpreter's build from first dialogue to first fully instrumented assignment (Zhao, 2026a).

\subsection{2.3 Translation as Human--Computer Interaction}

Translation studies has long treated the translator's work as a form of human--computer interaction, from the characterization of translation as interaction with tools that shape cognitive rhythms (O'Brien, 2012) to later arguments against antagonistic framings of human and machine capability in favor of human-centered augmentation (O'Brien, 2024). A framework paper in the present journal translates human-centered AI principles into design guidance for language technology along the whole pipeline (Briva-Iglesias \& O'Brien, 2026), and it addresses the designers of such systems. Within interpreting, a substantial computer-assisted interpreting literature examines tools designed for interpreters by developers outside the profession, including the foundational programmatic statements (Fantinuoli, 2018), the move toward AI-enhanced assistance (Fantinuoli, 2023), reviews of technology in interpreting practice (Chan, 2025), and experimental studies of assistance effects on terminology and performance (Chen \& Kruger, 2023; Prandi, 2023). The interpreter's own side of the building process had received little direct attention before the founding study, which documented, in a first-person and dialogue-level record, how an interpreter built a working tool for the interpreter's own booth (Zhao, 2026a). The present paper continues that record past its first deployment and asks what sustained growth means for the wider field.

\subsection{2.4 Competence, Literacy, and Assessment}

Competence research in translation and interpreting supplies the measurement vocabulary that Section 5 draws upon. Models of translation competence specify the component skills that training and certification address (PACTE, 2017), effort models of simultaneous interpreting describe the cognitive economies under which interpreters work (Gile, 2009), and quality models for live speech-to-text transfer provide criterion-referenced accuracy scoring (Romero-Fresco \& P\"ochhacker, 2017). Literacy research has widened the lens from skill with particular products toward general capabilities. The lineage runs from digital literacy (Gilster, 1997) through AI literacy frameworks defining what users should understand and be able to evaluate about AI systems (Long \& Magerko, 2020) to machine translation literacy addressing professional audiences (Bowker, 2020). Assessment research has begun to organize such competencies into holistic matrices spanning generic, domain-specific, and ethical dimensions (Knoth et al., 2024); the evaluation of practitioners together with the tools their practice has grown, and the institutional formats that could certify such ensembles, remain open in these traditions, and Section 5 develops three propositions in that direction.

Taken together, the four literatures locate the present study with some precision: user-innovation and end-user development research supplies the theory of users as innovators, conversational code generation supplies the newly lowered implementation barrier, translation studies supplies a high-stakes professional domain, and competence research supplies the measurement traditions that any assessment of grown tools must build on. A single longitudinal case in which all four meet has been missing, and the remainder of this paper provides one.

\section{3. The Paradigm of Practice-Grown Tools}

This section names and articulates the paradigm at the center of this paper. A practice-grown tool is a software artifact that a domain expert originates for use in the expert's own working practice, builds through conversation with code-generating AI systems rather than through conventional programming, and then develops in increments as real work exposes successive needs; the paradigm of practice-grown tools denotes the wider pattern of human--computer interaction that takes shape around such artifacts. Under this pattern the practitioner acts not only as a user of finished tools or as an informant to professional toolmakers, but as the leading party in a continuing cycle in which practice presents problems, the expert specifies and judges, AI systems implement, and the resulting tool feeds back into the shape of the practice itself. The founding study (Zhao, 2026a) established the feasibility of the first full turn of this cycle in conference interpreting. The present paper concerns what sustains and generalizes that cycle, and what it implies for how practitioners and their tools should be evaluated together. Table~\ref{tab:1} contrasts this pattern with the two established modes by which practitioners have acquired their software.

\begin{table}[htbp]
\centering
\caption{Three modes of acquiring professional software tools.}
\label{tab:1}
\small
\begin{tabularx}{\textwidth}{lXXX}
\toprule
\textbf{Dimension} & \textbf{Purchased tools} & \textbf{Customized tools (EUD)} & \textbf{Practice-grown tools} \\
\midrule
Who builds & Vendor developers & User or local tailor, within a vendor design & The domain expert, with AI implementation \\
Starting point & Finished product & Finished product plus tailoring hooks & Minimum viable tool derived from the expert's workflow \\
Growth driver & Vendor roadmap & Occasional tailoring by a minority of users & Recurrent breakdown and repair in real practice \\
Knowledge source & Market research and support channels & User feedback passed to designers & The expert's own articulated judgment \\
Failure and repair & Vendor support queue & Customization lost on update & Repaired by the builder in the same cycle \\
Computation & Vendor infrastructure & Vendor infrastructure & Bounded local hardware owned by the expert \\
Evaluation focus & Tool acceptance & Tool fit & Practitioner together with the tools at hand (Section 5) \\
\bottomrule
\end{tabularx}
\end{table}

\subsection{3.1 From Using Tools to Growing Them}

The intellectual lineage of this shift was surveyed in Section 2.1. User innovation established the demand, end-user development and meta-design supplied the ambition, and the toolkits literature worked out the economics. The recurring obstacle across that lineage was the cost of implementation; users could specify what they needed, yet building it remained expensive, so most user innovation stayed local, fragile, or dependent on professional intermediaries.

Conversational code generation has changed the terms of this long-standing obstacle in a specific way. When a practitioner can describe a desired behavior in professional language and receive working implementation from a coding agent, the skill barrier between knowing what a tool should do and being able to make it do so is substantially lowered (Karpathy, 2025; Sarkar \& Drosos, 2025). The change is already visible at the scale of whole communities, where public tooling for AI-assisted reverse engineering accumulated several hundred thousand users within two months of release and where the observed practice extends beyond imitation toward understanding and rebuilding the software on which the members' own workflows depend (morluto, 2026). Empirical study of design professionals points in the same direction, with code generation entering expert work that carries no formal programming training (Kobiella et al., 2026).

Two routes into tool creation deserve explicit distinction, as they encode different relations between the expert and what already exists.

In artifact-first reconstruction, the first of the two, the expert works from the outside in, starting with an existing finished product, disassembling it into an implementable understanding, and reassembling a variant suited to the needs of its reconstructors. Practice-first construction reverses the direction, starting from the expert's own workflow and growing a tool outward from a skeleton designed around that workflow. Case evidence in this paper, in which an interpreter and a translator each designed tool architectures around the stages of their own work, illustrates the second route. Seen together, the two routes are complementary, and their convergence suggests an emerging division of labor that this paper terms the skeleton-and-foliage division of labor. Professional judgment is irreplaceable exactly in designing the load-bearing skeleton, and that design therefore falls to the expert. AI generation can take the foliage, from user interface and import--export to file compatibility and platform integration, for that work consists more of labor than of judgment.

A pattern of this kind claims more than a technique does, and the claim deserves an explicit statement. Classic end-user development produced valued artifacts, yet these remained personal customizations, single adjustments made once by a skilled few and layered on top of finished products. Three features carry the present pattern beyond those bounded customizations. The tools serve the practitioner's own production work, on which professional output and livelihood depend, and they therefore inherit the reliability demands of that work. Growth is continuous and open-ended, sustained across versions, hardware, and model generations, where tailoring was typically episodic. And the expert leads the whole cycle in judgment, specifying, evaluating, and accepting, while the machine contributes implementation at scale. The combination earns the paradigm's name; no single ingredient does so on its own.

\subsection{3.2 Co-Evolution Through Breakdown and Repair}

\begin{figure}[htbp]
\centering
\includegraphics[width=0.9\textwidth]{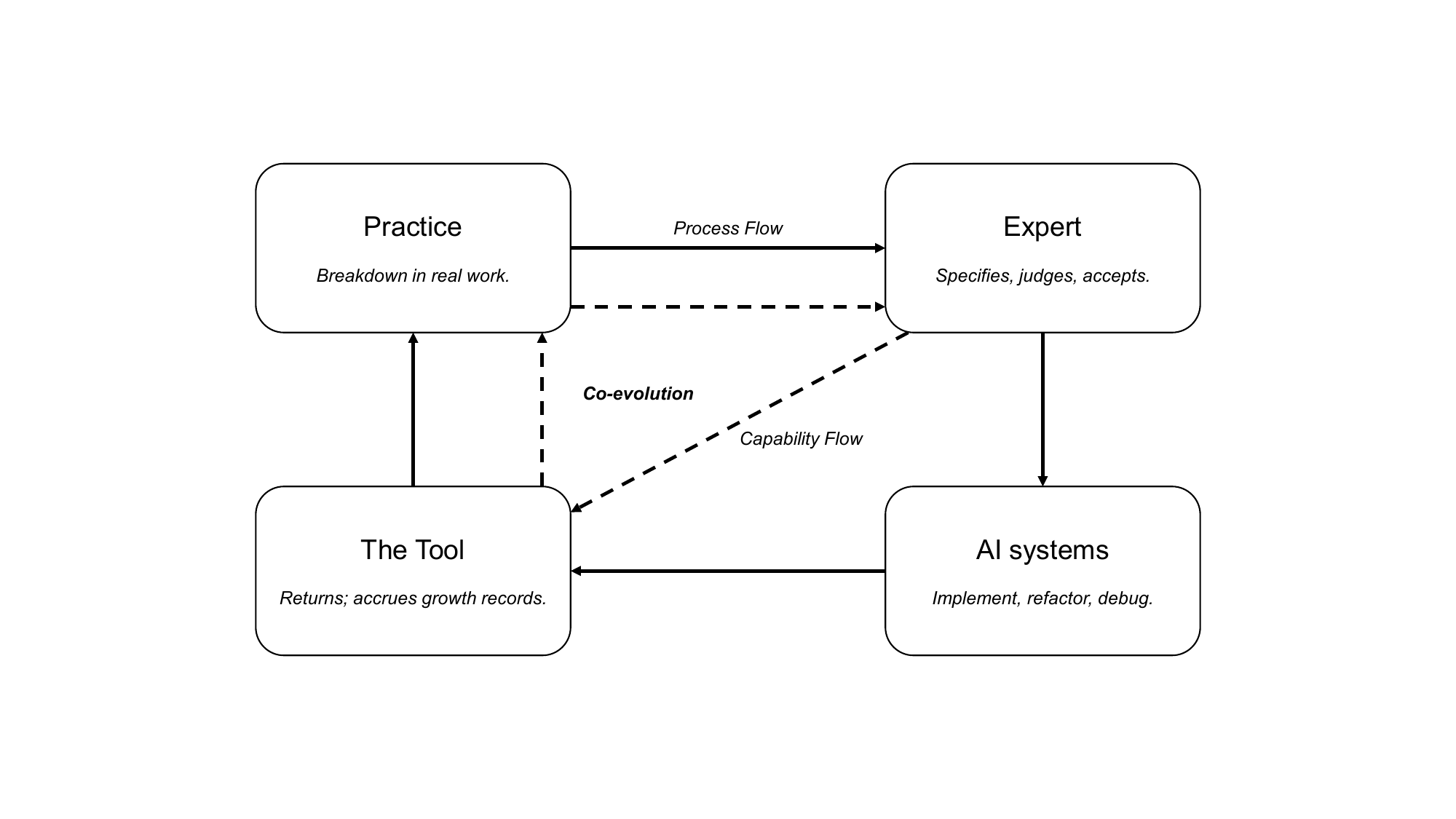}
\caption{The co-evolutionary loop of practice-grown tools. The outer ring (solid arrows) traces the process flow of breakdown and repair; the inner ring (dashed arrows) traces the capability flow through which the tool reshapes practice and the expert's judgment grows.}
\label{fig:1}
\end{figure}

A tool of this kind begins as a minimum viable tool (MVT), a term this paper adapts from the lean startup vocabulary of the minimum viable product (Ries, 2011) to name the smallest artifact that already does real work. From that starting point the tool grows through the loop sketched in Figure~\ref{fig:1} and documented in detail by the case in Section 4. Each iteration starts with a breakdown in real practice; the expert articulates what the practice actually requires; AI systems implement the specification; and the repaired tool returns to work until the next breakdown.

Each half of this loop can be anchored in an established theoretical tradition. The growth half follows the model of seeding, evolutionary growth, and reseeding, in which systems are delivered as seeds, evolve through use, and are periodically consolidated (Fischer, 1998). The skeleton of a practice-grown tool is precisely such a seed, and its foliage is the evolutionary growth that the seed was designed to invite. The trigger half follows the interactionist literature on repair. In that view, breakdown is not the failure of use but the moment at which the real relations between people and their technologies become visible and renegotiable, and repair work is knowledgeable, creative, and world-making rather than restorative (Jackson \& Kang, 2014; Jackson et al., 2012). On this reading, every row of the growth record assembled in Section 4 is a breakdown and repair episode, and the professional knowledge that the loop encodes becomes architecture exactly at the points where practice defeated the tool's previous form. A third anchor addresses what the growing relation does to the practitioner. The instrumental genesis tradition describes how an artifact becomes an instrument for a subject only through use, as usage schemes form around it and transform both the tool and its user (Rabardel \& Bourmaud, 2003). A practice-grown tool compresses this genesis, since its principal user is also its author, and each repair cycle re-articulates the instrument while the practice that formed it continues to move.

The version history of the case in Section 4 makes this rhythm visible in an unusually direct way. There, a version tag was placed not on the commit that introduced a long-planned feature but on the commit that repaired a failure observed during a real event, as if the version number itself recorded the moment at which the tool had survived practice.

\subsection{3.3 Bounded Local Computation as a Property of the Work Mode}

The third property concerns where the computation happens. In professional use, all of it runs on the practitioner's own portable hardware, with speech recognition and language models resident in local memory and with no dependence on network services. Boundedness here goes beyond being offline, since speech content stays on the practitioner's own disks by default, latencies are fixed by local hardware alone, and the failure surface is the machine in the room itself. The claim concerns the shape of the interaction between expert and machine, and the case evidence accordingly reports absolute processing speed only within each hardware generation. Confidential professional settings cannot expose speech content to external services (Fantinuoli, 2023), venues cannot be assumed to provide reliable connectivity, and an expert who builds a tool for daily work gains a specific kind of control over its behavior when the entire chain runs on hardware that the expert owns and administers. Human-centered AI has argued for high levels of meaningful human control over AI systems (Shneiderman, 2022). Local and bounded computation is one of the concrete forms such control takes for the expert who grows personal tools, and hardware constraints become design material for the co-evolutionary loop, as the case history shows when a change of machine became the occasion for a full re-architecture.

\subsection{3.4 An Assessment Corollary and the Boundaries of the Paradigm}

The fourth property follows as a corollary, and Section 5 develops it into three propositions. If tools now co-evolve with professional practice, then assessments of practitioners that strip the tools away measure a construct that practice has left behind, and the evaluation of a professional will increasingly have to reckon with the tools that the practice has grown.

The paradigm is offered as a candidate rather than a settled fact, and its boundaries should be stated plainly. The evidence base of this paper is one deeply documented case of growth over eight weeks in a single professional domain, together with a partial replication in a second tool built by the same practitioner. It is expected to apply where professional knowledge is dense and largely tacit, where existing tools leave visible gaps, where portable computation in the consumer range suffices for the task, and where conversational code generation has lowered the implementation barrier. It is not claimed for safety-critical components under formal certification regimes, nor for routine work whose demands are already well met by mass-market software. These boundaries are not concessions but specifications, for a candidate paradigm earns its generality precisely by declaring where it should not be expected to hold. The claim concerns the coherence of the pattern itself, while its prevalence across professions remains a matter for the replication studies anticipated in Section 6.

\section{4. Case: The Tool's Growth After the Founding Study}

\subsection{4.1 Data and Method}

The case evidence covers the interval between the initial commit on August 27, 2026 and October 22 of the same year, during which the tool passed through 62 commits and six tagged versions while serving real interpreting work. Three classes of evidence support the analysis: the version-control record provides the growth history, including commit messages that often state, in the practitioner's own words, what a change delivered or which event's feedback drove it; telemetry logs from 133 sessions, of which 84 contain captured speech, pair every recognized utterance with its recognition and translation timings and yield 20,707 recognized utterances in total; and synchronized audio recordings, held for instrumented sessions but neither analyzed nor distributed. The present analysis follows the founding study (Zhao, 2026a) and takes this eight-week record as its data, with all utterance-level statistics recomputed from the deposited dataset.

The method is a longitudinal single-case design in the analytic-generalization tradition and in the spirit of research in the wild, studying the tool where it actually works (Rogers \& Marshall, 2017). It rests on one researcher-practitioner in one professional domain, and the limitations that follow from this scope are taken up in Section 6. Where the argument attributes an effect to a specific development decision, the attribution is made only at points where configuration, hardware, and working day were held constant, and comparisons across hardware generations are avoided throughout.

\subsection{4.2 Starting Point}

The founding study established that a professional interpreter without programming training could build a working tool for the interpreter's own booth through a triadic co-creation workflow, and it reported the first fully instrumented assignment at a median two-leg translation latency of 0.94\,s across 971 utterances (Zhao, 2026a); the tool itself is released as open source under a citable archival version that carries the complete version history analyzed here (Zhao, 2026b). The version-control record begins with the stable pre-rewrite build committed on August 27, followed within days by the v0.2 rewrite, whose most consequential addition was the research logging chain itself. This instrumentation was therefore an early growth decision, taken so that the tool could serve as its own measurement device.

\subsection{4.3 Four Phases of Growth}

\begin{figure}[htbp]
\centering
\includegraphics[width=\textwidth]{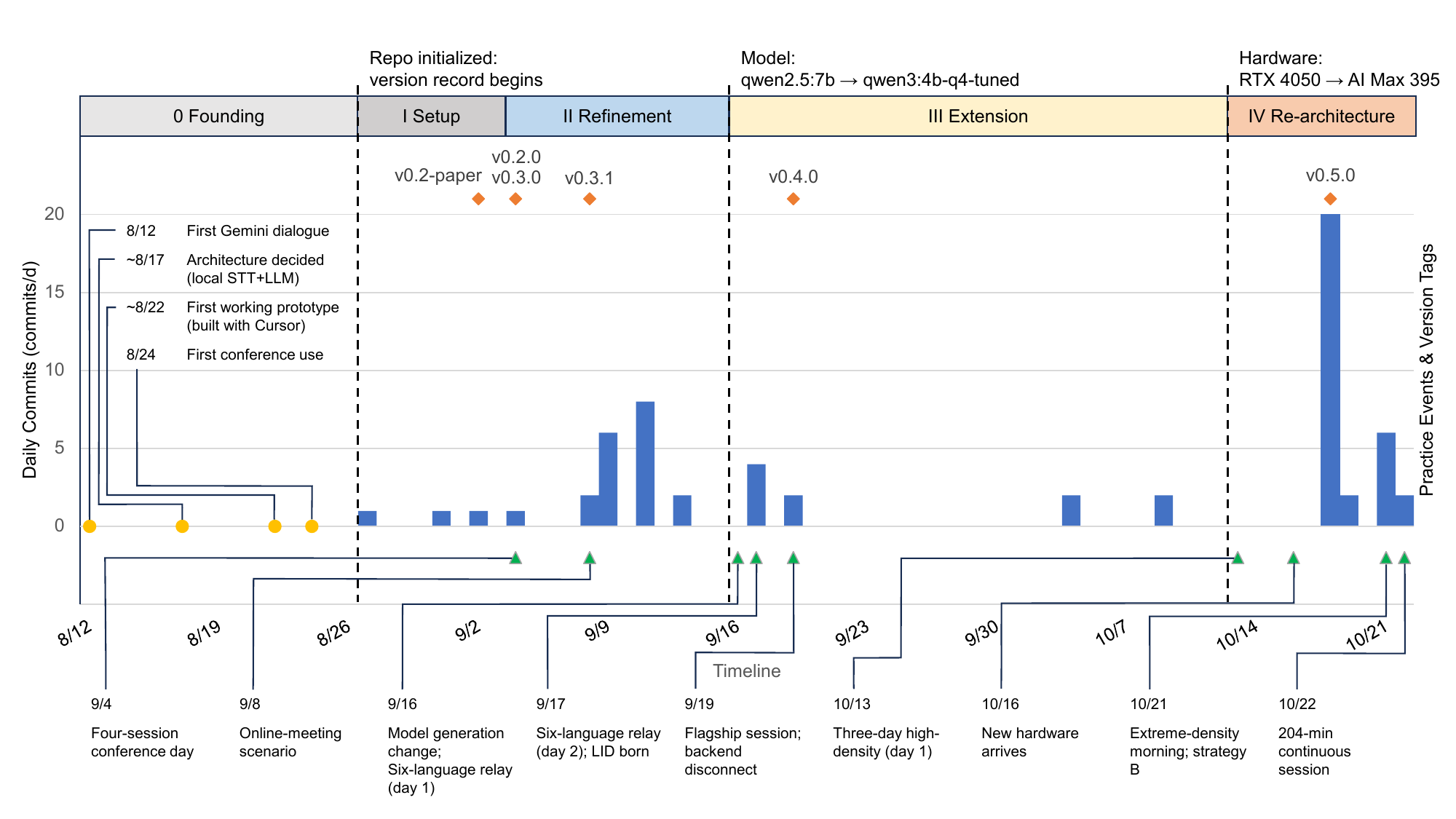}
\caption{Development timeline from August 12 to October 22, 2026. The hatched founding period (August 12--26) marks milestones from the initial build (Zhao, 2026a); the version-controlled record begins August 27. Daily commits, tagged versions, practice events, generation boundaries, and the four phases of Table~\ref{tab:2} appear as calendar layers.}
\label{fig:2}
\end{figure}

Read row by row, Table~\ref{tab:2} turns the growth record into a chain of breakdown and repair episodes in the sense of Section 3.2, and Figure~\ref{fig:2} lays the same chain on a calendar of commits, tags, practice events, and hardware generations. The chain falls into four phases, each with a characteristic trigger. The paragraphs that follow supply what neither display shows on its own, namely the mechanisms behind the rows.

\begin{table}[htbp]
\centering
\caption{Growth record as practice events and tool responses, grouped into four phases (August 27 to October 22, 2026; founding-era narrative in Zhao, 2026a).}
\label{tab:2}
\scriptsize
\begin{tabularx}{\textwidth}{llXXX}
\toprule
\textbf{Phase} & \textbf{Date} & \textbf{Practice event} & \textbf{Tool response} & \textbf{Observed effect} \\
\midrule
I & Aug 27--31 & Founding-era close-out (Zhao, 2026a) & v0.2 rewrite; logging, segmentation, backend & Tool becomes its own measurement device \\
II & Sep 4 & Four-session conference day & Same-day v0.3.0 subtitle upgrades & Logging enriched by that day's data \\
II & Sep 8 & Online-meeting scenario & v0.4.0 system-audio capture & New work mode supported \\
III & Sep 16 & Relay-work demands & Model replaced by smaller successor & Generation change post-dates founding telemetry \\
III & Sep 16--17 & Six-language relay forum; no LID & Three capability commits on Sep 17 & Capability born on the day its need appeared \\
III & Sep 19 & Flagship session; backend disconnects & v0.4.0 tag on repair commit & Version marks survival, not function \\
IV & Oct 13--15 & Three-day high-density forum & No changes triggered & 5,484 utterances; experiment becomes infrastructure \\
IV & Oct 16--18 & New hardware arrives & Migration and re-architecture; v0.5.0 & VAD rework; debug mode \\
IV & Oct 21 & Extreme-density morning; cap 41.8\% & Same-afternoon splitting strategy B & Cap share 23.0\%; see Table~\ref{tab:3} \\
IV & Oct 22 & 204-minute continuous session & Dedup gate upgraded; sleep protection & 1,097 utterances, zero failures \\
\bottomrule
\end{tabularx}
\end{table}

The second phase, in early September, refined the tool against feedback from conference work, with the four-session conference day of September 4 drawing a same-day release of subtitle upgrades and, four days later, a logging upgrade driven by that day's own data (Table~\ref{tab:2}).

The third phase was triggered by the demands of multilingual relay work. The same fortnight saw the translation stage itself change generation, as the locally quantized model of the founding study gave way to a smaller successor tuned for booth work; the migration post-dates every telemetry figure in the founding study and therefore belongs to the present record. At the publicly reported six-language relay forum of September 16 and 17, the telemetry still carried no language tags at all, and three capability commits landed on September 17 itself to add the missing identification capability, born on the day its need was discovered (Table~\ref{tab:2}).

The fourth phase began when new hardware arrived on October 16 and, over two days and some twenty commits, turned a machine change into the full re-architecture described in Section 3.3, reworked voice-activity detection included, released as v0.5.0. Its first real conference a day later carried the first language tags in the entire log, closing the loop that had opened at the relay forum. The three-day, high-density forum of October 13 to 15 had already passed without triggering a single change. On October 21, an extreme-density morning pushed the 15-second segment cap above 40 percent of utterances, and a punctuation-based splitting strategy deployed the same afternoon brought it down again; a 204-minute continuous session on October 22 then closed the record (Tables~\ref{tab:2} and~\ref{tab:3}).

\subsection{4.4 Outcomes Under Working Conditions}

\begin{table}[htbp]
\centering
\caption{Outcome evidence for four measurement sessions (end-to-end = configured silence confirmation + recognition + translation; $n$ = recognized utterances; latency distributions over matched utterances: 812, 1{,}086, 302, 628; cap share over all recognized utterances).}
\label{tab:3}
\small
\begin{tabular}{lrrrrrr}
\toprule
\textbf{Session} & $n$ & \textbf{Median} & $\leq$2\,s & $\leq$3\,s & \textbf{Cap} & \textbf{Rate} \\
 & & \textbf{E2E} & & & \textbf{share} & \textbf{(CJK/s)} \\
\midrule
Sep 17 relay & 812 & 1{,}513\,ms & 92.0\% & 99.8\% & 0\% & --- \\
Oct 21 AM & 1{,}140 & 2{,}461\,ms & 34.5\% & 84.2\% & 41.8\% & 4.28 \\
Oct 21 PM & 387 & 1{,}548\,ms & 78.1\% & 99.7\% & 23.0\% & 4.05 \\
Oct 22 & 1{,}097 & 2{,}017\,ms & 49.0\% & 97.1\% & 46.6\% & 3.87 \\
\bottomrule
\end{tabular}
\end{table}

Table~\ref{tab:3} reports the three-tier outcome evidence for four measurement sessions, using the end-to-end measure defined as the configured silence confirmation plus the recognition and translation legs for matched utterances, a stricter definition than the two-leg measure of the founding study, and the two calibrations should not be compared as though they were the same quantity.

\subsection{4.5 A Second Tool Grown by the Same Practice}

\begin{figure}[htbp]
\centering
\includegraphics[width=0.95\textwidth]{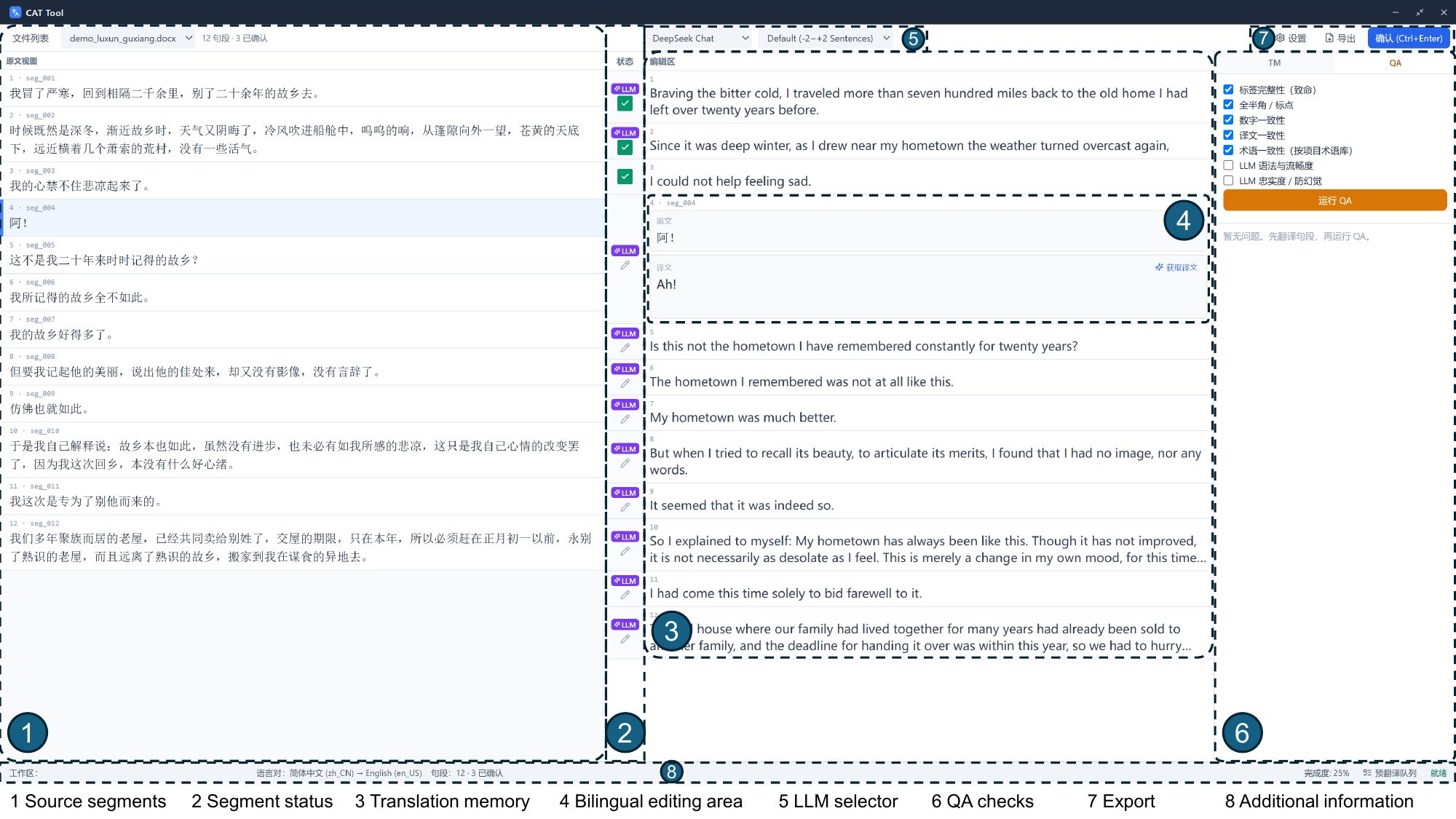}
\caption{The second tool at its current early version: the bilingual editing view on a demonstration project. Interface labels are in the practitioner's working language; numbered callouts mark the main interface modules.}
\label{fig:3}
\end{figure}

A partial replication exists in a second tool begun by the same practitioner for written translation. Its architecture was designed around the stages of a translator's workflow, with modules for import and parsing, translation-memory and terminology assets, bilingual editing, quality checking, and export, arranged around a service layer for local language models, and it reuses the technical approach of the first tool. At its current early version, the core chain operates on test projects while the outer layers of file-format support and platform integration remain to be grown. The example matters less for its present completeness than for its direction, since the architecture was drawn from the inside out, from the practitioner's own workflow, in the manner that Section 3 called practice-first construction, and it suggests that the growing of tools, once learned, transfers across the tasks of a practice. Figure~\ref{fig:3} shows the second tool at its current early version, operating on a demonstration project.

\section{5. Propositions: Toward Tool-Inclusive Assessment}

The three properties developed in Section 3 describe how practice-grown tools come into being and continue to develop. The present section draws out what follows from them for the assessment of practitioners. A profession that acquires its tools in this way can no longer evaluate its practitioners as if the tools did not exist. The argument is offered as three propositions, and each is stated as a direction for design and research.

\subsection{5.1 From Tool Proficiency to Tool-Making Literacy}

Professional education and certification have generally treated technological competence as a matter of tool proficiency, widely understood as the skilled operation of finished products selected by others. Literacy frameworks widened this view considerably, from digital literacy through machine translation literacy to general conceptions of AI literacy that cover understanding, applying, and critically evaluating AI systems (Bowker, 2020; Gilster, 1997; Long \& Magerko, 2020). Practice-grown tools shift the center of the construct once more. What distinguishes a practitioner who can grow tools is not familiarity with any particular product but the ability to convert professional knowledge into working specifications, documented in the founding study as a practiced triad of specifying, judging, and accepting (Zhao, 2026a). Tool-making literacy, as this paper uses the term, names the disposition and skill to perform that conversion, to orchestrate AI implementation toward the practitioner's own requirements, and to maintain the result under real working conditions.

The first proposition nonetheless faces a substantial objection concerning equality of access. A technology that promises wider access to tool creation carries a high threshold of its own, since conversational code generation still presumes hardware, environments, and habits of specification that most practitioners have yet to acquire. This accessibility paradox, in which an inclusion-enhancing technology is itself difficult to enter, recalls the second-level digital divide, in which access to a technology and the effective skill to use it come apart (Hargittai, 2002). The paradox is resolved not by the technology itself but by institutions. Layered supply, anticipated in the tailoring tradition (MacLean et al., 1990), lets the tools grown by a few experts serve many colleagues, and formal training converts the threshold into a curriculum. Literacy, on this reading, is the institutional mechanism by which a candidate paradigm becomes progressively inclusive. The assessment designs of the two remaining propositions are where that mechanism becomes concrete.

\subsection{5.2 The Tool-at-Hand as the Unit of Assessment}

With the tool now participating constitutively in professional performance, the unit of assessment should be the practitioner together with the tools actually at hand, not the practitioner alone under artificially stripped conditions. The recognition that language work is itself a form of human--computer interaction predates the present generation of tools (O'Brien, 2012), and ensemble assessment follows that recognition through to its measurement consequences. Interpreting research already offers part of the needed measurement tradition, since ear--voice span, delivery proportions, and error-based accuracy measures have long described interpreter performance, and criterion-referenced accuracy scoring for live speech-to-text transfer was itself developed for human and machine working together (Gile, 2009; Romero-Fresco \& P\"ochhacker, 2017). The evaluation of human--machine ensembles is likewise an open measurement question across human--AI teaming research, where frameworks for reporting joint human and machine contribution are under active development (National Academies of Sciences, Engineering, and Medicine, 2022). Validated workload instruments add the subjective dimension that ensemble performance is likely to change first. Translation competence research has built detailed models of what a translator must master (PACTE, 2017). The corresponding move for written translation runs from text-level toward project-level assessment, as a translator's tool capability expresses itself in the setup, the terminology assets, and the process quality of a whole engagement. Recording the tool configuration alongside the human scores keeps the object of record the ensemble and results interpretable across configurations. A practice-grown tool adds one further possibility to this record, since its version history documents how the practitioner's knowledge entered the tool, episode by episode. An auditable growth record is stronger evidence of tool-making literacy than any self-report or product demonstration could provide.

\subsection{5.3 Compute-Standardized Tool-Creation Assessment}

Assessment regimes have so far taken two main routes with AI, and neither yet addresses the ensemble. Knowledge-oriented examinations in law and accounting prohibit AI use during testing, and the exclusion fits their construct, for what such examinations certify is knowledge held by the individual. Performance-oriented certification in the language professions tests live professional work, yet it does so under deliberately unaided conditions, even though the work it simulates is now commonly aided. Written translation examinations have long permitted dictionaries, so the exclusion of digital assistance is a matter of degree rather than of kind, while interpreting examinations remain fully closed. Other performance professions have meanwhile attached certification to the equipment itself, as aviation does when a license is rated for a particular aircraft type (14 C.F.R. \S\,61.31, 2026). Robotic surgery has built validated, platform-specific curricula that anchor its training and credentialing pathways (Martin et al., 2021). What results is a specific gap, since the language professions assess unaided performance inside an aided practice, and no existing format yet evaluates the act of growing or holding the tools that the practice now includes.

The third proposition, compute-standardized tool-creation assessment, describes a format that addresses this gap while keeping comparison fair. All candidates receive identical computing terminals under a published, uniform specification of compute, together with the same set of task texts. Within a unified model environment provided by the organizers, they create, configure, or adapt their tools on the spot through conversational code generation, and the grown tools then face the same performance scenarios. Scoring weighs the skeleton, understood as the design decisions that encode domain knowledge, more heavily than the foliage that candidates remain free to delegate. Competition aeromodeling illustrates the regulatory logic at work, since equal batteries, mass limits, organizer-supplied flight instruments, and shared tasks channel creativity into design and away from equipment budgets; separate design-report judging rewards the engineering reasoning behind the airframe (American Institute of Aeronautics and Astronautics, 2025; China Universities Aircraft Design Competition, 2026).

Two validity threats are foreseeable, and both are answerable within the format itself. Prepared tools could be smuggled in under the guise of on-site adaptation; from-scratch construction within the unified environment, with published component libraries, would largely foreclose that route. Generative AI could equally be accused of doing the real work, and skeleton-weighted scoring answers that objection directly, certifying the professional judgment embodied in the architecture rather than the volume of generated code. The proposition is stated as a design hypothesis for research, and the micro-studies needed to calibrate it are themselves instances of the assessment research that the paradigm calls for.

\section{6. Discussion}

For the vendors of professional software, the case suggests a shift in where innovation originates and how it travels. The growth record shows that the features a practitioner grows fastest are precisely those that respond to needs too idiosyncratic for a vendor roadmap to capture, and a mature division of labor is therefore easy to imagine, in which standardized cores remain vendor territory while the long tail of professional constraint is grown by practitioners themselves. The paradigm even implies a product strategy in Fischer's (1998) vocabulary, since a vendor that ships seeds rather than finished products, designed to be grown by their users, converts its most demanding customers into co-developers. Whether vendors move in this direction is an empirical question the present case cannot settle, but the founding study and the growth record together make the demand side visible in a way that market research rarely achieves.

The reach of the paradigm also extends beyond software into the arrangements of hardware that professional work runs on. At the six-language relay forum of Section 4, the interpreting infrastructure itself was a recomposition of commodity parts, with interpreters working on an industry-standard console whose translated audio was fed to a portable wireless tour-guide transmitter and from there reached handheld receivers across the hall, an arrangement the venue's engineers had assembled for reasons of coverage, usability, and cost. The hall was too wide for the fixed infrared radiators to cover without additional cabling that the client declined; receivers preset to one channel per language spared audience members the mis-tuning that multi-channel sets invite; and the whole chain cost a fraction of a proprietary installation. That chain carried the tool through the console's own output, and the tool lived inside it. The frontier the paradigm points toward is the devolution of such recomposition capability to practitioners, as conversational agents extend from writing new software toward understanding existing hardware, and simplified chains of this kind could in principle bring simultaneous interpreting within reach of settings and regions that proprietary chains price out. These prospects are stated at the level of potential, with the boundaries that honest judgment requires, since reverse engineering carries legal limits, reliability ceilings matter in high-stakes settings, and the appropriate first deployments are low-stakes and community-scale. At the community level, the same agent-based reverse engineering is already used to revive and maintain software and devices past their official end of life, against the grain of planned obsolescence (Blevis, 2007; Slade, 2006).

The development carries three risks that deserve to be named and answered. Generated code has documented quality and security weaknesses, and a practice-grown tool inherits them; extending the practitioner's judgment into testing and maintenance is therefore part of the literacy that the curricula of Section 5 propose. Responsibility needs new descriptions, since a tool grown by its user enters professional workflows without vendor warranty, and questions of liability and credit will travel with every ensemble. And the success of one capable practitioner can be mistaken for a general capability. The accessibility discussion of Section 5 is designed to forestall exactly this over-reading.

The limitations of the evidence are structural and should be stated without hedging. It rests on a single expert-led study in one professional domain and one language market, and the author's combined role as developer, user, and analyst is a reflexivity that full process transparency can bound but not eliminate. Telemetry covers the tool alone, so aided performance on delivery measures remains unmeasured. Across the growth record's two hardware generations, the boundaries are declared rather than controlled. Appropriate next studies are comparative performance work on aided interpreting, replication by practitioners in other domains, and the assessment micro-studies that would calibrate the third proposition. The deposited dataset and the open repository exist precisely to make those studies cheaper than this first one was.

\section{7. Conclusion}

This paper examined whether a route that a single interpreter proved workable can become a paradigm, and pursued the question in four moves. The first move named the paradigm of practice-grown tools and gave it four properties, with co-evolution through breakdown and repair at the center and a skeleton-and-foliage division of labor as its emerging ideal. The second supplied a longitudinal growth record of eight weeks of versioned development under real working conditions, in which reliability held under sustained load and a same-day strategy deployment cut the median delay by over a third and more than doubled the share of utterances inside two seconds, under near-identical delivery conditions. The third argued that assessment should follow the ensemble, and it proposed tool-making literacy, the tool-at-hand as the unit of evaluation, and a compute-standardized format for certifying creation itself. The fourth marked the paradigm's boundaries. The candidate applies where expertise is dense, tools leave gaps, and portable computation suffices, and it declines to apply elsewhere. The tools have begun to grow, and the open question is whether the institutions that educate and certify practitioners can grow with them.

\section*{References}

\begin{small}
\setlength{\parskip}{0.3em}
\setlength{\parindent}{-1.5em}
\setlength{\leftskip}{1.5em}
\hspace*{-1.5em}American Institute of Aeronautics and Astronautics. (2025). \emph{Design/Build/Fly competition rules, 2025--2026}. \url{https://aiaa.org/wp-content/uploads/2025/08/2026-DBF-Rules.pdf}

Baldwin, C. Y., \& von Hippel, E. (2011). Modeling a paradigm shift: From producer innovation to user and open collaborative innovation. \emph{Organization Science, 22}(6), 1399--1417. \url{https://doi.org/10.1287/orsc.1100.0618}

Blevis, E. (2007). Sustainable interaction design: Invention \& disposal, renewal \& reuse. In \emph{Proceedings of the SIGCHI Conference on Human Factors in Computing Systems} (pp. 503--512). ACM. \url{https://doi.org/10.1145/1240624.1240705}

Bowker, L. (2020). Machine translation literacy instruction for international business students and business English instructors. \emph{Journal of Business \& Finance Librarianship, 25}(1--2), 25--43. \url{https://doi.org/10.1080/08963568.2020.1794739}

Briva-Iglesias, V., \& O'Brien, S. (2026). Human-centered AI language technology (HCAILT): An empathetic design framework for reliable, safe and trustworthy multilingual communication. \emph{International Journal of Human--Computer Interaction, 42}(19), 16632--16646. \url{https://doi.org/10.1080/10447318.2026.2622588}

Chan, V. (2025). Impact of technology on interpreting practice: A review of studies on technology and interpreting practice from 2013 to 2024. \emph{Interactive Technology and Smart Education, 22}(1), 81--102. \url{https://doi.org/10.1108/ITSE-02-2024-0042}

Chen, S., \& Kruger, J.-L. (2023). The effectiveness of computer-assisted interpreting: A preliminary study based on English--Chinese consecutive interpreting. \emph{Translation and Interpreting Studies, 18}(3), 399--420. \url{https://doi.org/10.1075/tis.21036.che}

China Universities Aircraft Design Competition. (2026). \emph{Competition rules} (Version 20260330). Chinese Society of Aeronautics and Astronautics.

Collard, C., \& Defrancq, B. (2019). Predictors of ear--voice span, a corpus-based study with special reference to sex. \emph{Perspectives, 27}(3), 431--454. \url{https://doi.org/10.1080/0907676X.2018.1553199}

Davis, F. D. (1989). Perceived usefulness, perceived ease of use, and user acceptance of information technology. \emph{MIS Quarterly, 13}(3), 319--340. \url{https://doi.org/10.2307/249008}

Fantinuoli, C. (2018). Interpreting and technology: The upcoming turn. In C. Fantinuoli (Ed.), \emph{Interpreting and technology} (pp. 1--12). Language Science Press. \url{https://doi.org/10.5281/zenodo.1493281}

Fantinuoli, C. (2023). Towards AI-enhanced computer-assisted interpreting. In G. Corpas Pastor \& B. Defrancq (Eds.), \emph{Interpreting technologies -- Current and future trends} (pp. 46--71). John Benjamins. \url{https://doi.org/10.1075/ivitra.37.03fan}

Fischer, G. (1998). Seeding, evolutionary growth and reseeding: Constructing, capturing and evolving knowledge in domain-oriented design environments. \emph{Automated Software Engineering, 5}(4), 447--464.

Fischer, G., Giaccardi, E., Ye, Y., Sutcliffe, A. G., \& Mehandjiev, N. (2004). Meta-design: A manifesto for end-user development. \emph{Communications of the ACM, 47}(9), 33--37.

Ge, Y., Mei, L., Duan, Z., Li, T., Zheng, Y., Wang, Y., Wang, L., Yao, J., Liu, T., Cai, Y., Bi, B., Guo, F., Guo, J., Liu, S., \& Cheng, X. (2025). A survey of vibe coding with large language models. \emph{arXiv}. \url{https://arxiv.org/abs/2510.12399}

Gile, D. (2009). \emph{Basic concepts and models for interpreter and translator training} (Rev. ed.). John Benjamins.

Gilster, P. (1997). \emph{Digital literacy}. Wiley.

Hargittai, E. (2002). Second-level digital divide: Differences in people's online skills. \emph{First Monday, 7}(4). \url{https://doi.org/10.5210/fm.v7i4.942}

Huynh, P., \& Rivero, A. (2026). A scoping review of vibe coding in healthcare across clinical, administrative, and research domains [Preprint]. \emph{Research Square}. \url{https://doi.org/10.21203/rs.3.rs-10529517}

Jackson, S. J., \& Kang, L. (2014). Breakdown, obsolescence and reuse: HCI and the art of repair. In \emph{Proceedings of the SIGCHI Conference on Human Factors in Computing Systems} (pp. 449--458). ACM. \url{https://doi.org/10.1145/2556288.2557332}

Jackson, S. J., Pompe, A., \& Krieshok, G. (2012). Repair worlds: Maintenance, repair, and ICT for development in rural Namibia. In \emph{Proceedings of the ACM 2012 Conference on Computer Supported Cooperative Work}. ACM. \url{https://doi.org/10.1145/2145204.2145224}

Karpathy, A. (2025, February 2). There's a new kind of coding I call ``vibe coding''\ldots{} [Post]. X.

Knoth, N., Decker, M., Laupichler, M. C., Pinski, M., Buchholtz, N., Bata, K., \& Schultz, B. (2024). Developing a holistic AI literacy assessment matrix: Bridging generic, domain-specific, and ethical competencies. \emph{Computers and Education Open, 6}, 100177. \url{https://doi.org/10.1016/j.caeo.2024.100177}

Kobiella, C., Breidenstein, D., \& Schmidt, A. (2026). From throw-away to takeaway: How GenAI and vibe coding accelerate prototyping across technical skill levels. In \emph{Proceedings of the 2026 CHI Conference on Human Factors in Computing Systems}. ACM. \url{https://doi.org/10.1145/3772318.3790757}

Lieberman, H., Patern\`o, F., Klann, M., \& Wulf, V. (2006). End-user development: An emerging paradigm in human-computer interaction. In H. Lieberman, F. Patern\`o, \& V. Wulf (Eds.), \emph{End user development} (pp. 1--8). Springer.

Long, D., \& Magerko, B. (2020). What is AI literacy? Competencies and design considerations. In \emph{Proceedings of the 2020 CHI Conference on Human Factors in Computing Systems}. ACM. \url{https://doi.org/10.1145/3313831.3376727}

MacLean, A., Carter, K., L\"ovstr\"om, L., \& Moran, T. (1990). User-tailorable systems. In \emph{Proceedings of the SIGCHI Conference on Human Factors in Computing Systems} (pp. 175--182). ACM. \url{https://doi.org/10.1145/97243.97271}

Martin, J. R., Stefanidis, D., Dorin, R. P., Goh, A. C., Satava, R. M., \& Levy, J. S. (2021). Demonstrating the effectiveness of the fundamentals of robotic surgery (FRS) curriculum on the RobotiX Mentor virtual reality simulation platform. \emph{Journal of Robotic Surgery, 15}(2), 187--193. \url{https://doi.org/10.1007/s11701-020-01085-4}

morluto. (2026). REA: Reverse engineer anything with agents [Computer software]. GitHub. \url{https://github.com/morluto/rea}

National Academies of Sciences, Engineering, and Medicine. (2022). \emph{Human-AI teaming: State-of-the-art and research recommendations}. National Academies Press. \url{https://doi.org/10.17226/26355}

Nielsen, J. (1993). \emph{Usability engineering}. Academic Press.

O'Brien, S. (2012). Translation as human--computer interaction. \emph{Translation Spaces, 1}(1), 101--122. \url{https://doi.org/10.1075/ts.1.05obr}

O'Brien, S. (2024). Human-centered augmented translation: Against antagonistic dualisms. \emph{Perspectives, 32}(3), 391--406. \url{https://doi.org/10.1080/0907676X.2023.2247423}

PACTE. (2017). \emph{Researching translation competence by PACTE group}. John Benjamins.

Pimenova, V., Fakhoury, S., Bird, C., Storey, M.-A., \& Endres, M. (2026). Good vibrations? A qualitative study of co-creation, communication, flow, and trust in vibe coding. In \emph{Proceedings of the 2026 CHI Conference on Human Factors in Computing Systems}. ACM. \url{https://doi.org/10.1145/3772318.2509.12491}

Prandi, B. (2023). \emph{Computer-assisted simultaneous interpreting: A cognitive-experimental study on terminology}. Language Science Press.

Rabardel, P., \& Bourmaud, G. (2003). From computer to instrument system: A developmental perspective. \emph{Interacting with Computers, 15}(5), 665--691.

Ries, E. (2011). \emph{The lean startup}. Crown Business.

Rogers, Y., \& Marshall, P. (2017). \emph{Research in the wild}. Morgan \& Claypool.

Romero-Fresco, P., \& P\"ochhacker, F. (2017). Quality assessment in interlingual live subtitling: The NTR model. \emph{Linguistica Antverpiensia, 16}, 149--167.

Sarkar, A., \& Drosos, I. (2025). Vibe coding: Programming through conversation with artificial intelligence. In \emph{Proceedings of the 36th Annual Conference of the Psychology of Programming Interest Group}. \url{https://arxiv.org/abs/2506.23253}

Shneiderman, B. (2022). \emph{Human-centered AI}. Oxford University Press.

Slade, G. (2006). \emph{Made to break: Technology and obsolescence in America}. Harvard University Press.

Type rating requirements, additional training, and authorization requirements, 14 C.F.R. \S\,61.31 (2026). \url{https://www.ecfr.gov/current/title-14/chapter-I/subchapter-D/part-61/section-61.31}

von Hippel, E. (1986). Lead users: A source of novel product concepts. \emph{Management Science, 32}(7), 791--805. \url{https://doi.org/10.1287/mnsc.32.7.791}

von Hippel, E. (2005). \emph{Democratizing innovation}. MIT Press.

von Hippel, E., \& Katz, R. (2002). Shifting innovation to users via toolkits. \emph{Management Science, 48}(7), 821--833. \url{https://doi.org/10.1287/mnsc.48.7.821.2817}

Zhao, Z. (2026a). The interpreter as developer: An autoethnographic design study of building a computer-assisted interpreting tool through human--LLM co-creation. \emph{arXiv}. \url{https://arxiv.org/abs/2610.16460}

Zhao, Z. (2026b). whisper\_interpreter (Version v0.2-paper) [Computer software]. Zenodo. \url{https://doi.org/10.5281/zenodo.22255011}

\end{small}

\end{document}